\documentclass[aps,prb,twocolumn,showpacs,preprintnumbers,amsmath,amssymb,superscriptaddress]{revtex4}%

\usepackage{graphicx}
\usepackage{dcolumn}
\usepackage{bm}
\usepackage{color}

\begin{document} 

\preprint{preprint(\today)}

\title{Probing the pairing symmetry in the over-doped Fe-based superconductor  Ba$_{0.35}$Rb$_ {0.65}$Fe$_{2}$As$_{2}$ as a function of hydrostatic pressure}
\author{Z.~Guguchia}
\email{zurab.guguchia@psi.ch} \affiliation{Laboratory for Muon Spin Spectroscopy, Paul Scherrer Institute, CH-5232
Villigen PSI, Switzerland}

\author{R.~Khasanov}
\affiliation{Laboratory for Muon Spin Spectroscopy, Paul Scherrer Institute, CH-5232
Villigen PSI, Switzerland}

\author{Z.~Bukowski}
\affiliation{Institute of Low Temperature and Structure Research, Polish Academy of Sciences, 50-422 Wroclaw, Poland}

\author{ F.~von Rohr}
\affiliation{Physik-Institut der Universit\"{a}t Z\"{u}rich, Winterthurerstrasse 190,
CH-8057 Z\"{u}rich, Switzerland}

\author{M.~Medarde}
\affiliation{Laboratory for Developments and Methods, Paul Scherrer Institute, CH-5232 Villigen PSI, Switzerland}


\author{P.K.~Biswas}
\affiliation{Laboratory for Muon Spin Spectroscopy, Paul Scherrer Institute, CH-5232
Villigen PSI, Switzerland}

\author{H.~Luetkens}
\affiliation{Laboratory for Muon Spin Spectroscopy, Paul Scherrer Institute, CH-5232
Villigen PSI, Switzerland}

\author{A.~Amato}
\affiliation{Laboratory for Muon Spin Spectroscopy, Paul Scherrer Institute, CH-5232
Villigen PSI, Switzerland}

\author{E.~Morenzoni}
\affiliation{Laboratory for Muon Spin Spectroscopy, Paul Scherrer Institute, CH-5232
Villigen PSI, Switzerland}

\begin{abstract}

  We report muon spin rotation experiments on the magnetic penetration depth $\lambda$ and the temperature dependence of $\lambda$$^{-2}$ in the  over-doped Fe-based high-temperature superconductor (Fe-HTS) Ba$_{1-x}$Rb$_ {x}$Fe$_{2}$As$_{2}$ ($x$ = 0.65) studied at ambient and under hydrostatic pressures up to $p$ = 2.3 GPa. We find that in this system $\lambda$$^{-2}$($T$) is best described by $d$-wave scenario. This is in contrast to the case of the optimally doped $x$ = 0.35 system which is known to be a nodeless $s^{+-}$-wave superconductor. This suggests that the doping induces the change of the pairing symmetry from $s^{+-}$ to $d$-wave in Ba$_{1-x}$Rb$_ {x}$Fe$_{2}$As$_{2}$. In addition, we find that the $d$-wave order parameter is robust against pressure, suggesting that $d$ is the common and dominant pairing symmetry in over-doped  
Ba$_{1-x}$Rb$_{x}$Fe$_{2}$As$_{2}$. Application of pressure of $p$ = 2.3 GPa causes a decrease of $\lambda$(0) by less than 5 ${\%}$, while at optimal doping $x$ = 0.35 a significant decrease of $\lambda$(0) was reported. The superconducting transition temperature $T_{c}$ as well as the gap to $T_{\rm c}$ ratio 2${\Delta}$/$k_{\rm B}$$T_{\rm c}$ show only a modest decrease with pressure. By combining the present data with those previously obtained for optimally doped system $x$ = 0.35 and for the end member $x$ = 1 we conclude that the SC gap symmetry as well as the pressure effects on the SC quantities strongly depend on the Rb doping level. These results are discussed in the light of the putative  Lifshitz transition, i.e., a disappearance of the electron pockets in the Fermi surface of Ba$_{1-x}$Rb$_ {x}$Fe$_{2}$As$_{2}$  upon hole doping. 


\end{abstract}

\pacs{74.20.Mn, 74.25.Ha, 74.70.Xa, 76.75.+i, 62.50.-p}

\maketitle

\section{Introduction}

 The family of unconventional superconductors has grown considerably over the last couple of decades and now includes cuprates\cite{Bednorz}, heavy-fermions \cite{Steglich}, organic superconductors \cite{Jerome} and most recently also iron pnictides \cite{Kamihara08,Takahashi}. They all share a similar phase diagram \cite{Uemura2009,Chu-2009}. Superconductivity emerges through doping or applied pressure when the competing magnetic state is suppressed. 
Even after more than 20 years of intensive research the superconducting (SC) pairing mechanism is still not understood for the above mentioned compounds \cite{ZhaoG}. 
To understand it, it is instructive to study the symmetry and structure of the SC gap. 
A significant experimental and theoretical effort has concentrated on studies of this issue in Fe-HTS's. However, there is no consensus on a universal gap structure and the relevance for the particular gap symmetry for high-temperature superconductivity in iron-based high temperature superconductors (Fe-HTS's), which are the first non-cuprate materials exhibiting superconductivity at relatively high temperatures. 
In contrast to cuprates, where the SC gap symmetry is universal the gap symmetry and/or structure of the Fe-HTS's can be quite different from material to material.
For instance, nodeless isotropic gap distributions were observed in optimally doped Ba$_{1-x}$K$_{x}$Fe$_{2}$As$_{2}$, Ba$_{1-x}$Rb$_{x}$Fe$_{2}$As$_{2}$ and BaFe$_{2-x}$Ni$_{x}$As$_{2}$ as well as in BaFe$_{2-x}$Co$_{x}$As$_{2}$, K$_{x}$Fe$_{2-y}$Se$_{2}$ and 
FeTe$_{1-x}$Se$_{x}$ \cite{Ding, KhasanovN, GuguchiaN, TerashimaN, ZhangY, MiaoN,Abdel-Hafiez,BiswasPRB}. 
Signatures of nodal SC gaps were reported in LaOFeP, LiFeP, KFe$_{2}$As$_{2}$,
BaFe2(As$_{1-x}$P$_{x}$)$_{2}$, BaFe$_{2-x}$Ru$_{x}$As$_{2}$ as well as in over-doped 
Ba$_{1-x}$K$_{x}$Fe$_{2}$As$_{2}$ and BaFe$_{2-x}$Ni$_{x}$As$_{2}$ 
\cite{FlechterN, HashimotoN, YamashitaN, NakaiN, HashimotoK, DongN, QiuN, SongN, ZhangN,Abdel-Hafiez}. 

\begin{figure}[b!]
\includegraphics[width=1.07\linewidth]{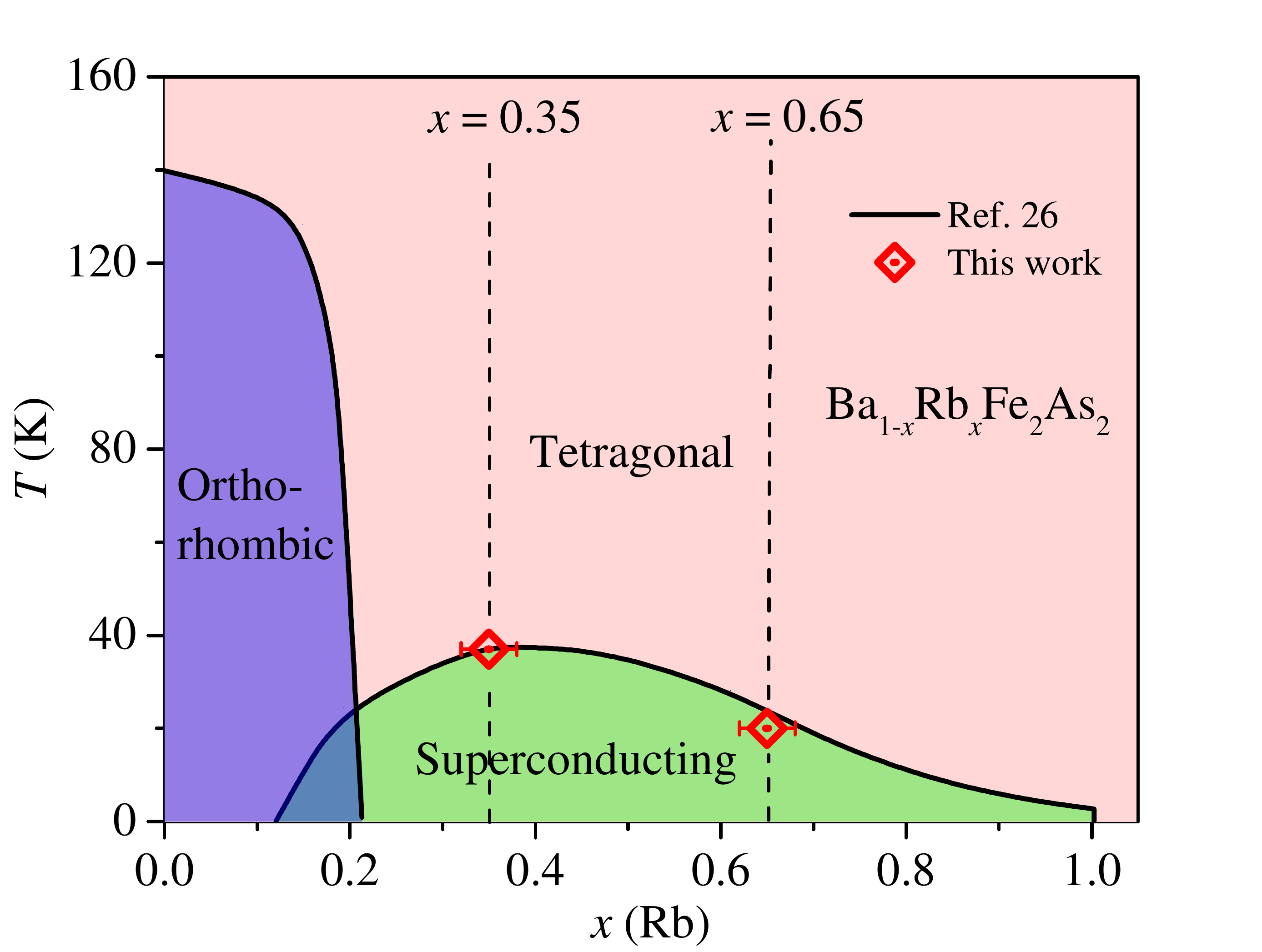}
\vspace{-0.7cm}
\caption{ (Color online) Schematic phase diagram of Ba$_{1-x}$Rb$_{x}$Fe$_{2}$As$_{2}$ (after Ref.~26). The open symbols represent the values of the SC transition temperature obtained in this work.
The dashed line mark the doping level for our sample.}
\label{fig1}
\end{figure}

 We note that an important feature of Ba$_{1-x}$Rb$_{x}$Fe$_{2}$As$_{2}$ and the related system 
Ba$_{1-x}$K$_{x}$Fe$_{2}$As$_{2}$  is that the SC phase exists over a wide range of Rb and K-concentration, respectively, namely from $x$ = 0.2 to $x$ = 1.
For clarity, the schematic phase diagram for Ba$_{1-x}$Rb$_{x}$Fe$_{2}$As$_{2}$, taken from Ref.~26 
is shown in Fig.~1. The data points obtained in the present work are also shown. 
It was found that this phase diagram is very similar to the thoroughly studied system  
Ba$_{1-x}$K$_{x}$Fe$_{2}$As$_{2}$ \cite{Peschke}.
The particularly interesting observation in Ba$_{1-x}$K$_{x}$Fe$_{2}$As$_{2}$ is the systematic doping evolution of 
the nodal structure for heavy hole-doping \cite{Rheid-2012,Hirano2012}. 
At around  optimal doping  $x$ = 0.4, where $T_{\rm c}$ has a maximum value of 38 K, many experiments revealed the occurrence of multiple isotropic SC gaps. A sign changing $s_{±}$-wave state which is mediated by spin fluctuations has been invoked to explain some of the experimental results. However, this state is expected to be very fragile
to the presence of nonmagnetic impurities, while Fe-HTS's are experimentally
known to be robust against nonmagnetic impurities. A no-sign changing
$s_{++}$-wave state mediated by orbital fluctuations and which
is robust against nonmagnetic impurities is another possible
candidate for the pairing in this system. Hence, the SC pairing symmetry in the optimum region is
still an open question. However, there is consensus that
the SC gap structure itself is fully gapped for optimally doped samples of Ba$_{1-x}$K$_{x}$Fe$_{2}$As$_{2}$.
It is  interesting that in this system, the crossover from nodeless to nodal SC state occurs at
$x$ ${\sim}$ 0.8. These changes were related to a Lifshitz transition, reflecting the
disappearance of the electron pockets in the Fermi surface (FS), at similar K-doping levels \cite{Hirano2012}
(unlike optimally doped Ba$_{0.6}$K$_{0.4}$Fe$_{2}$As$_{2}$ 
which has both electron and hole-FSs, only hole-FSs were found in the extremely hole-doped KFe$_{2}$As$_{2}$ \cite{Sato}). The nodeless SC gaps were also observed in Ba$_{1-x}$Rb$_{x}$Fe$_{2}$As$_{2}$ at optimal doping $x$ = 0.3, 0.35, 0.4 from the temperature dependence of the magnetic penetration depth ${\lambda}$ by means of muon-spin rotation (${\mu}$SR) \cite{GuguchiaN}.  ${\mu}$SR experiments performed on
polycrystalline samples of extremely hole-doped RbFe$_{2}$As$_{2}$ also suggested the presence of two isotropic $s$-wave gaps \cite{Shermadini,Shermadinipressure}. However, recent specific heat and thermal conductivity measurements on single crystals of RbFe$_{2}$As$_{2}$ and on the related compound CsFe$_{2}$As$_{2}$ provided evidence for nodal SC gap in these materials \cite{Zhang2015,Wang2013,Hong2013}. This suggests that the crossover from nodeless to nodal state upon hole doping should also be present  in Ba$_{1-x}$Rb$_{x}$Fe$_{2}$As$_{2}$.  In this regard it is important to study the SC gap symmetry in over-doped 
Ba$_{1-x}$Rb$_{x}$Fe$_{2}$As$_{2}$.

 Besides doping another important tuning parameter is the hydrostatic pressure which leads to new and in some materials very exotic physical properties, pressure induced phase transitions as well as changes of the characteristic SC or magnetic quantities \cite{KlaussPressure,KhasanovPressure,GuguchiaPressure,Taillefer,GuguchiaSubmitted,Prando}. 
In KFe$_{2}$As$_{2}$, a change of the SC pairing symmetry by hydrostatic
pressure has been proposed, based on the $V$-shaped pressure
dependence of $T_{{\rm c}}$ \cite{Taillefer}. Recently, we have shown unambiguous evidence for the appearance of SC nodes
in optimally-doped Ba$_{1-x}$Rb$_{x}$Fe$_{2}$As$_{2}$ upon applied
pressure, consistent with a change from a nodeless $s^{+-}$-wave state to a $d$-wave state \cite{GuguchiaSubmitted}.  
Interestingly, the theoretical calculations \cite{Kuroki,Graser10,Thomale11,Maiti11,Khodas12,Fernandes13,FernandezN} as well as Raman experiments \cite{raman_mode,Bohm} revealed a sub-dominant $d$-wave state close in energy to the dominant $s^{+-}$ state. It seems that pressure affects this intricate balance, and tip the balance in favor of the $d$-wave state. Besides the appearance of nodes with pressure in  optimally-doped Ba$_{1-x}$Rb$_{x}$Fe$_{2}$As$_{2}$, another interesting observation was a strong decrease of the magnetic penetration depth ${\lambda}$ with pressure \cite{GuguchiaSubmitted}. In contrast, in the end member compound RbFe$_{2}$As$_{2}$ an increase of ${\lambda}$ and no change of the gap symmetry was found up to $p$ = 1.1 GPa \cite{Shermadinipressure}. Thus, it is important to study the pressure effects on the SC properties of over-doped 
Ba$_{1-x}$Rb$_{x}$Fe$_{2}$As$_{2}$ system in order to have a picture about the pressure effects on different regions of the phase diagram.

  In the following we report on ${\mu}$SR studies of  the temperature dependence of the penetration depth ${\lambda}$
in over-doped Ba$_{0.35}$Rb$_{0.65}$Fe$_{2}$As$_{2}$ at ambient and under hydrostatic pressures up to $p$ = 2.3 GPa. 
These results suggest the $d$-wave superconductivity in this system, which is distinctly different from the nodeless gap found at optimal doping. The $d$-wave order parameter symmetry is preserved under pressure. 
The SC transition temperature $T_{c}$, the value of the $d$-wave gap as well as the zero-temperature value of the magnetic penetration depth ${\lambda}$(0) show only a modest decrease with pressure. We compare the present pressure data with the previous results of optimally doped  Ba$_{0.65}$Rb$_{0.35}$Fe$_{2}$As$_{2}$ \cite{GuguchiaSubmitted} and the end member RbFe$_{2}$As$_{2}$  \cite{Shermadinipressure} and discuss the combined results in the light of the possible Lifshitz transition in Ba$_{1-x}$Rb$_{x}$Fe$_{2}$As$_{2}$ induced by hole doping.

\begin{figure}[b!]
\includegraphics[width=1.0\linewidth]{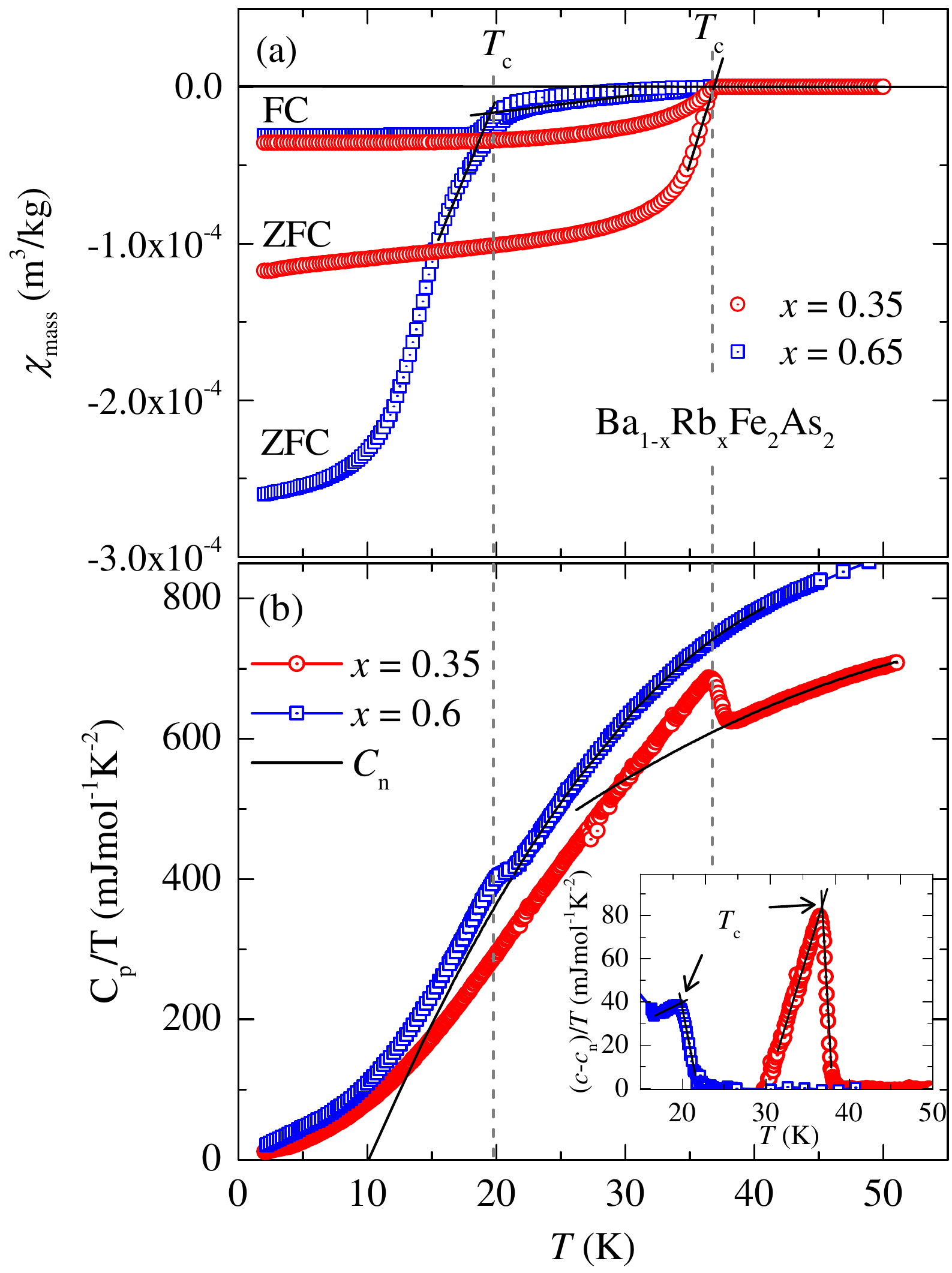}
\caption{ (Color online) Temperature dependence of the zero-field cooled (ZFC) and field-cooled (FC) susceptibility
obtained in an applied magnetic field of $\mu_{0}$$H$ = 2 mT for Ba$_{1-x}$Rb$_{x}$Fe$_{2}$As$_{2}$ ($x$ = 0.35, 0.65). (b) The specific heat $C_{\rm p}$/$T$ as a function of temperature for  Ba$_{1-x}$Rb$_{x}$Fe$_{2}$As$_{2}$ ($x$ = 0.35, 0.65). The dashed lines denote the superconducting transition temperatures $T_{c}$ for both samples. The solid line is the fitted normal state contribution $C_{\rm n}$. The inset shows the specific heat with $C_{\rm n}$ subtracted for different values of $x$.}
\label{fig1}
\end{figure}
\begin{figure}[b!]
\includegraphics[width=1.0\linewidth]{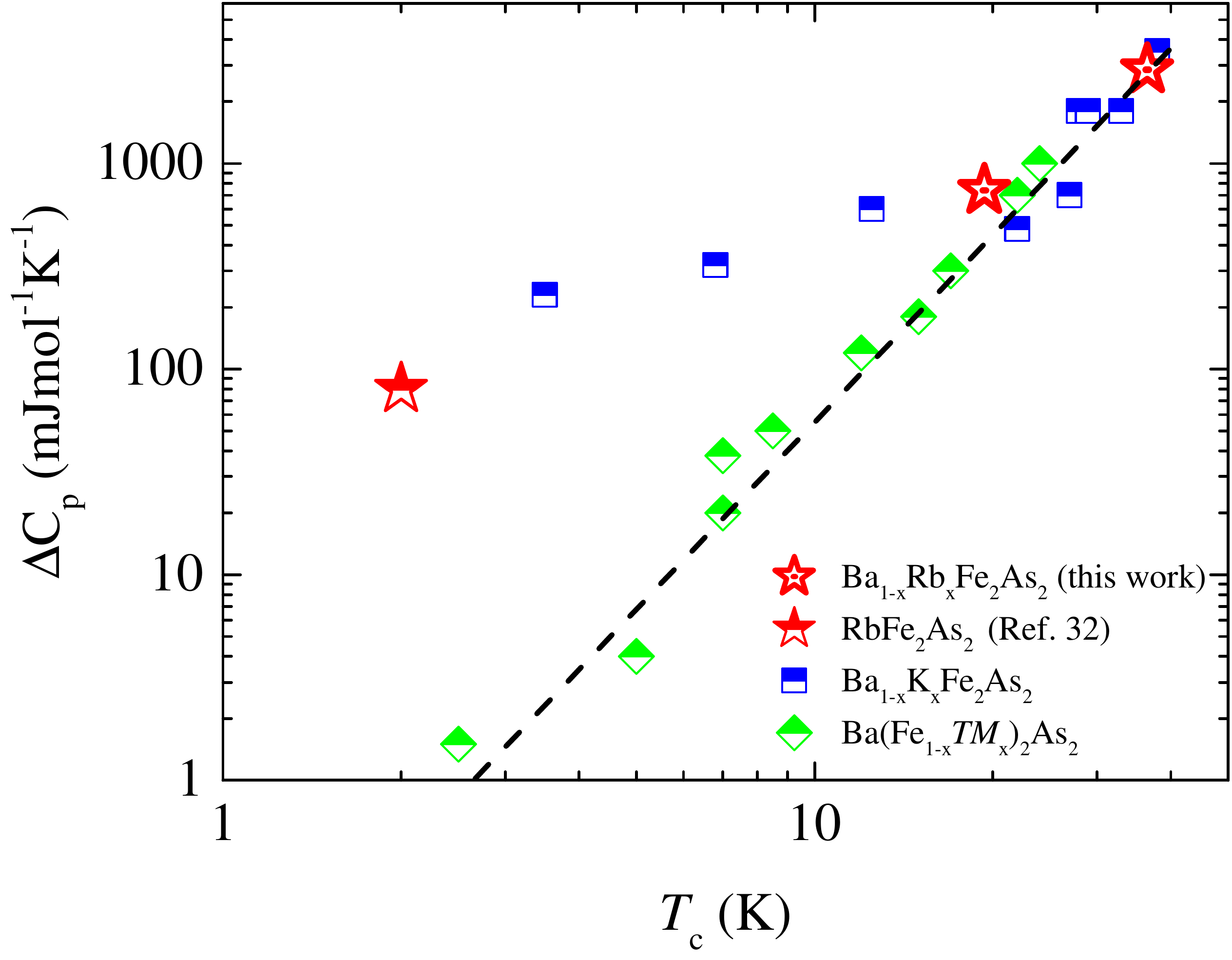}
\caption{ (Color online) $\Delta$$C_{\rm p}$ at the superconducting transition vs
$T_{\rm c}$ for Ba$_{1-x}$Rb$_{x}$Fe$_{2}$As$_{2}$ ($x$ = 0.35, 0.65), plotted together with literature
data for some Fe-based SC materials, belonging to so-called '122'  family (after Ref.~53). The data point for $x$ = 0.35 is taken from Ref.~39. 
The line corresponds to  $\Delta$$C_{\rm p}$ ${\propto}$ $T^{3}$.}
\label{fig1}
\end{figure}

\section{EXPERIMENTAL DETAILS }
 Polycrystalline samples of Ba$_{1-x}$Rb$_{x}$Fe$_{2}$As$_{2}$ ($x$ = 0.35, 0.65) were prepared in evacuated quartz ampoules 
by a solid state reaction method. Fe$_{2}$As, BaAs, and RbAs were obtained by 
reacting high purity As (99.999 $\%$), Fe (99.9$\%$), Ba (99.9$\%$) and Rb (99.95$\%$) 
at 800 $^{\circ}$C, 650 $^{\circ}$C and 500~$^{\circ}$C, respectively.
Using stoichiometric amounts of BaAs or RbAs and Fe$_{2}$As the terminal compounds BaFe$_{2}$As$_{2}$ 
and RbFe$_{2}$As$_{2}$ were synthesized at 950 $^{\circ}$C and 650 $^{\circ}$C, respectively. 
Finally, the samples of Ba$_{1-x}$Rb$_{x}$Fe$_{2}$As$_{2}$ with $x$ = 0.35, 0.65 were prepared from appropriate 
amounts of single-phase BaFe$_{2}$As$_{2}$ and RbFe$_{2}$As$_{2}$. The components were mixed, pressed into pellets, 
placed into alumina crucibles and annealed for 100 hours at 650 $^{\circ}$C with one 
intermittent grinding. Magnetization, specific heat, powder X-ray diffraction, and ${\mu}$SR experiments were performed on samples from the same batch. 
This allows to study the SC properties and the pressure effects in similar samples and to make direct comparison between the various properties in optimally and over-doped
Ba$_{1-x}$Rb$_{x}$Fe$_{2}$As$_{2}$. Powder X-ray diffraction analysis revealed that the synthesized samples 
are single phase materials. The magnetization measurements were performed with a commercial SQUID magnetometer
($Quantum~Design$ MPMS-XL). The specific heat measurements (relaxor type calorimeter
Physical Properties Measurements System $Quantum~Design$) were performed in zero field.
Zero-field (ZF) and transverse-field (TF) ${\mu}$SR 
experiments were performed at the ${\pi}$M3 beamline of the Paul Scherrer
Institute (Villigen, Switzerland), using the general purpose instrument (GPS). The sample was mounted inside of a
gas-flow $^4$He cryostat on a sample holder with a standard veto setup providing essentially a 
background free $\mu$SR signal. ${\mu}$SR experiments under various applied pressures were performed at the $\mu$E1 beamline of PSI, using the dedicated GPD spectrometer.
Pressures up to 2.3 GPa were generated in a double wall piston-cylinder
type of cell made out of MP35N material, especially designed to perform ${\mu}$SR experiments under
pressure \cite{MaisuradzePRB}. As a pressure transmitting medium Daphne oil was used. The pressure was measured
by tracking the SC transition of a very small indium plate by AC susceptibility.
All TF experiments were carried out after a field-cooling procedure. 
The ${\mu}$SR time spectra were analyzed using the free software package MUSRFIT \cite{Bastian}. 


\section{RESULTS AND DISCUSSION}
\subsection{Magnetization and specific heat capacity experiments}

 The temperature dependence of zero field-cooled (ZFC) and field-cooled (FC) diamagnetic susceptibility  measured in a magnetic field of $\mu_{\rm 0}$$H$ = 1 mT for Ba$_{1-x}$Rb$_{x}$Fe$_{2}$As$_{2}$ ($x$ = 0.35, 0.65) is shown in Fig. 2(a). The SC transition temperature $T_{\rm c}$ for $x$ = 0.35 is determined from the intercept of the linearly extrapolated zero-field cooled (ZFC) magnetization curve with 
$\chi_{mass}$ = 0 line and it is found to be $T_{\rm c}$ = 37 K. For the sample $x$ = 0.6, $T_{\rm c}$ = 20 K was found, using the intercept
of linear extrapolations above and below $T_{\rm c}$, due to a small fraction with higher $T_{\rm c}$. Temperature-dependent heat capacity data for both samples plotted as $C_{\rm p}$/$T$ vs $T$ are shown in Fig.~2(b). The jumps associated with the SC transitions are clearly seen for both concentrations. 
The observed strong diamagnetic response and the specific heat jumps at $T_{\rm c}$  provide solid evidence for bulk superconductivity in both compounds. To quantify the jump in specific heat, the anomaly at the transition has been isolated from the phonon dominated background by subtracting a second order polynomial $C_{\rm p,n}$ fitted above $T_{\rm c}$ and extrapolated to lower temperature \cite{Walsmley}. The quantity ($C_{\rm p}$ - $C_{\rm p,n}$)/$T$ is presented as a function of temperature in the inset of Fig.~2(b). 
Although there is some uncertainty in using this procedure over an extended temperature range, the lack of appreciable thermal SC fluctuations, as evidenced by the mean-field-like form of the anomaly, means that there is very little uncertainty in the size of ${\Delta}C_{\rm p}$. It is evident from the inset of Fig.~2(b) that the size of the anomaly ${\Delta}C_{\rm p}$/$T_{\rm c}$  depends very strongly on $x$ and $T_{\rm c}$. A strong increase in ${\Delta}C_{\rm p}$ with $T_{\rm c}$ has been observed previously in many '122' Fe-HTS's \cite{Budko1,Budko2,Budko3}. Bud'ko $et.~al.$ found that in many '122' Fe-HTS's the specific heat jump 
${\Delta}C_{\rm p}$ at $T_{\rm c}$ follow the empirical trend, the so-called BNC scaling \cite{Budko1,Budko2,Budko3} ${\Delta}C_{\rm p}$  ${\propto}$ $T^{3}$. This has been interpreted as either originating from quantum critically or from strong impurity pair breaking. A violation of the BNC scaling was observed for Ba$_{1-x}$K$_{x}$Fe$_{2}$As$_{2}$ with $x$ ${\textgreater}$ 0.7, for which a change of the SC gap symmetry/structure was found \cite{Budko2,Budko3}. The specific heat jump data for the  Ba$_{1-x}$Rb$_{x}$Fe$_{2}$As$_{2}$ samples obtained in this work and one for the end member RbFe$_{2}$As$_{2}$ \cite{Zhang2015} are added in Fig.~3, to the BNC plot taken from Ref.~\cite{Budko3}. The point for the optimally doped sample lies perfectly on the BNC line. On the other hand, the data point for $x$ = 0.65 sample is slightly off from it and the point for the end-compound RbFe$_{2}$As$_{2}$ clearly deviates from this scaling.
This indicates that the heavily over-doped Ba$_{1-x}$Rb$_{x}$Fe$_{2}$As$_{2}$ shows a deviation from the BNC scaling, 
similar to that observed in related Ba$_{1-x}$K$_{x}$Fe$_{2}$As$_{2}$. 
Hence, one expects significant changes in the nature of the SC state in the over-doped 
Ba$_{1-x}$Rb$_{x}$Fe$_{2}$As$_{2}$.

\subsection{${\mu}$SR experiments}
\begin{figure}[b!]
\includegraphics[width=0.75\linewidth]{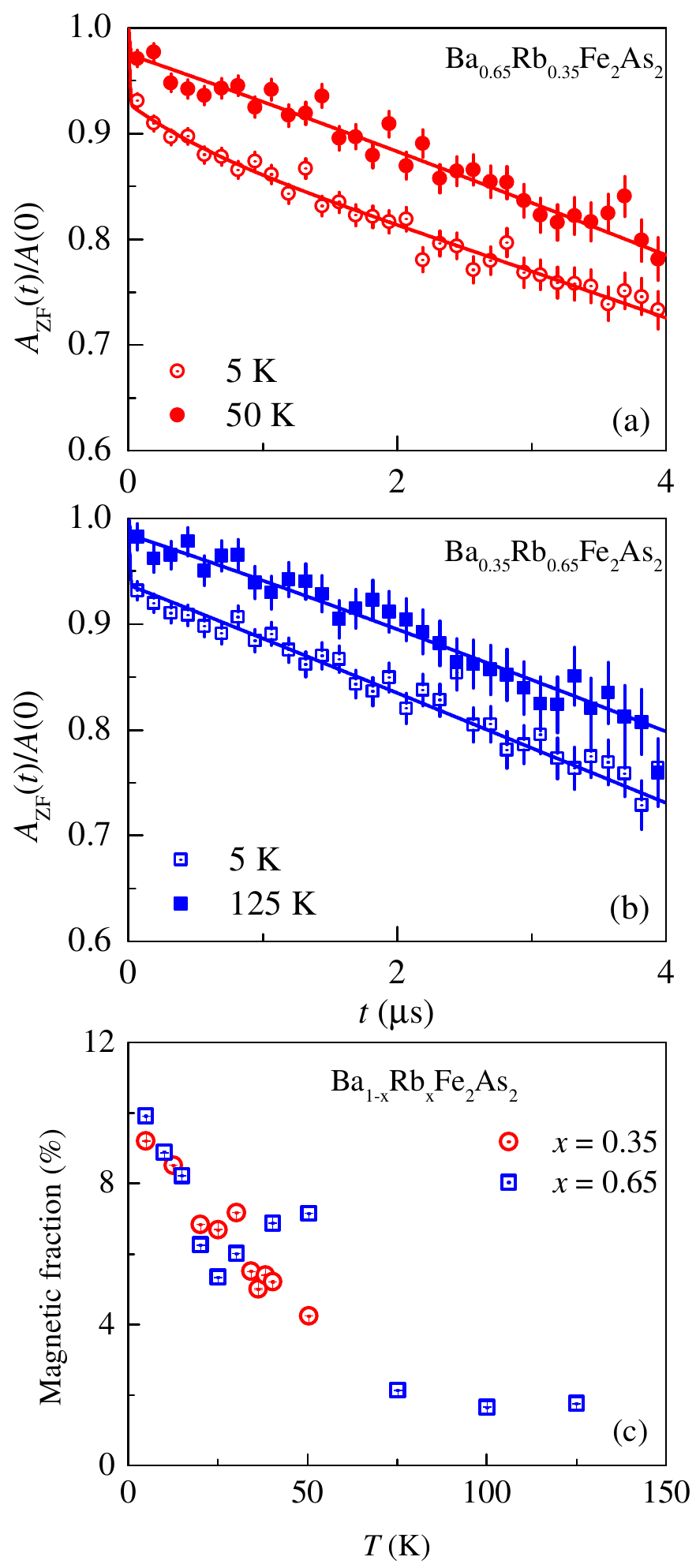}
\caption{(Color online) ZF-${\mu}$SR time spectra for Ba$_{0.65}$Rb$_{0.35}$Fe$_{2}$As$_{2}$ (a) 
and  Ba$_{0.35}$Rb$_{0.65}$Fe$_{2}$As$_{2}$ (b) recorded above and below $T_{\rm c}$. The line represents the fit to the data by means of Eq.~1.
(c) Temperature dependence of the magnetic fraction in Ba$_{0.65}$Rb$_{0.35}$Fe$_{2}$As$_{2}$ and Ba$_{0.35}$Rb$_{0.65}$Fe$_{2}$As$_{2}$, extracted from the ZF-${\mu}$SR experiments. The error bars represent the s.d. of the fit parameters and they are smaller than the symbols.}
\label{fig7}
\end{figure}
\subsubsection{Zero-field and transverse-field ${\mu}$SR on Ba$_{0.65}$Rb$_{0.35}$Fe$_{2}$As$_{2}$  and Ba$_{0.35}$Rb$_{0.65}$Fe$_{2}$As$_{2}$ at ambient pressure}

  It is well known that undoped BaFe$_{2}$As$_{2}$ is not superconducting
at ambient pressure and undergoes a spin-density wave (SDW) transition of the Fe-moments far above $T_{\rm c}$.\cite{Huang} The SC state can be achieved either under pressure \cite{Torikachvili,Miclea} or by appropriate
charge carrier doping \cite{Zhao} of the parent compounds, leading to a suppression of the SDW state.
Our first task was to check whether magnetism is present in the samples.
Therefore, we have carried out ZF-${\mu}$SR experiments above and below $T_{\rm c}$ in Ba$_{0.65}$Rb$_{0.35}$Fe$_{2}$As$_{2}$ and Ba$_{0.35}$Rb$_{0.65}$Fe$_{2}$As$_{2}$.
As an example ZF-${\mu}$SR spectra obtained above and below $T_{\rm c}$ are shown in Fig.~4a and b, respectively. There are no precession signals, indicating the absence of long-range magnetic order. On the other hand, we observed a significant drop of the asymmetry, taking place within  0.2 ${\mu}$s. This is caused by the presence of diluted Fe moments as discussed in previous ${\mu}$SR studies \cite{Khasanov}.  
In order to check the size of the magnetic fraction, the ZF-${\mu}$SR data were analyzed by the following function:

\begin{figure}[b!]
\includegraphics[width=1.0\linewidth]{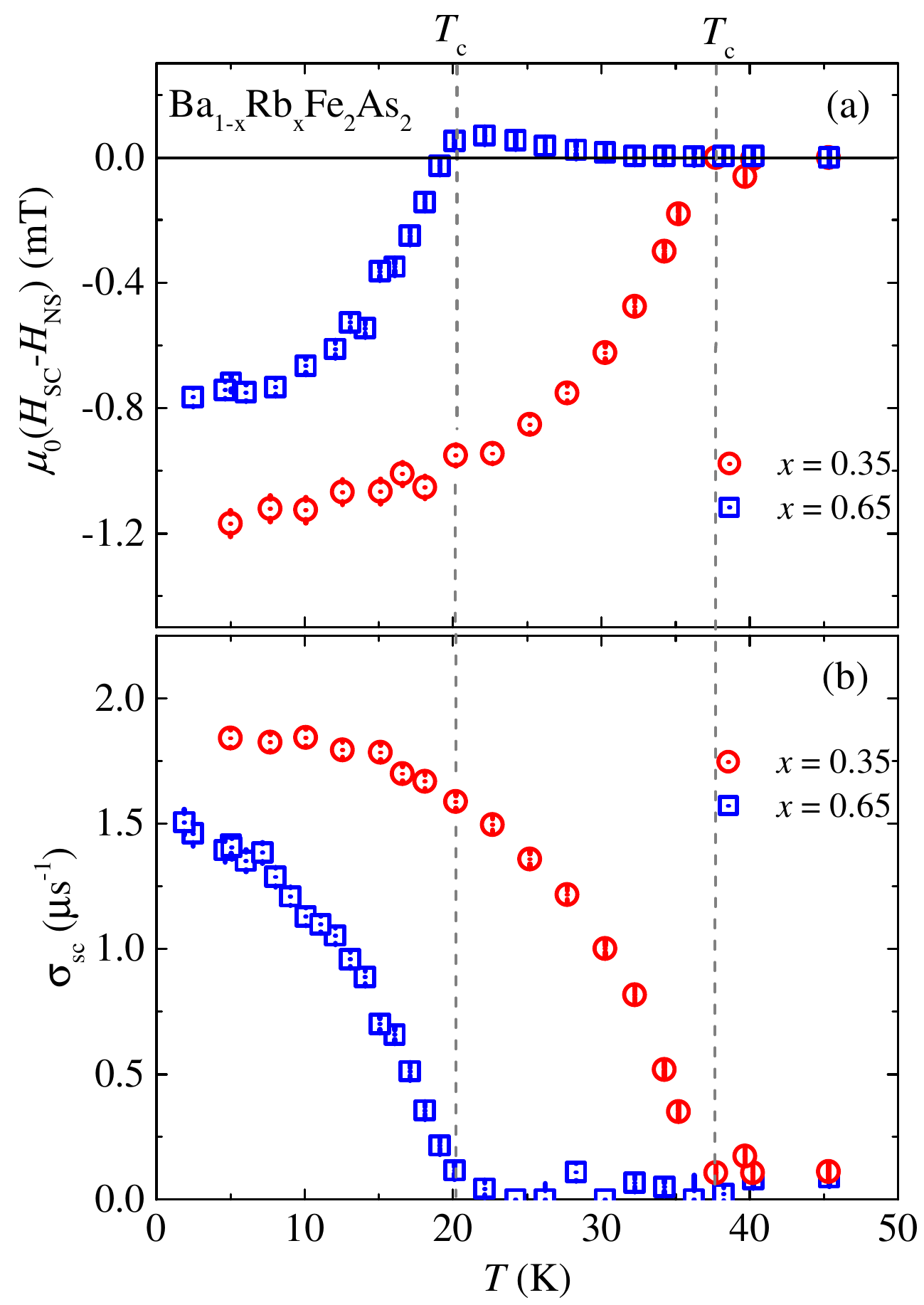}
\caption{ (Color online) (a) Temperature dependence of the difference
between the internal field ${\mu}_{\rm 0}$$H_{\rm SC}$ measured in the SC state 
and the one measured in the normal state ${\mu}_{\rm 0}$$H_{\rm NS}$ at $T$ = 50~K  for Ba$_{1-x}$Rb$_{x}$Fe$_{2}$As$_{2}$ ($x = 0.35$ and 0.65). 
(b) Temperature dependence of the  superconducting muon spin depolarization rate
${\sigma}_{\rm sc}$ measured in an applied magnetic field of ${\mu}_{\rm 0}H = 0.05$~T for both samples. }
\label{fig2}
\end{figure}
\begin{figure}[t!]
\includegraphics[width=1\linewidth]{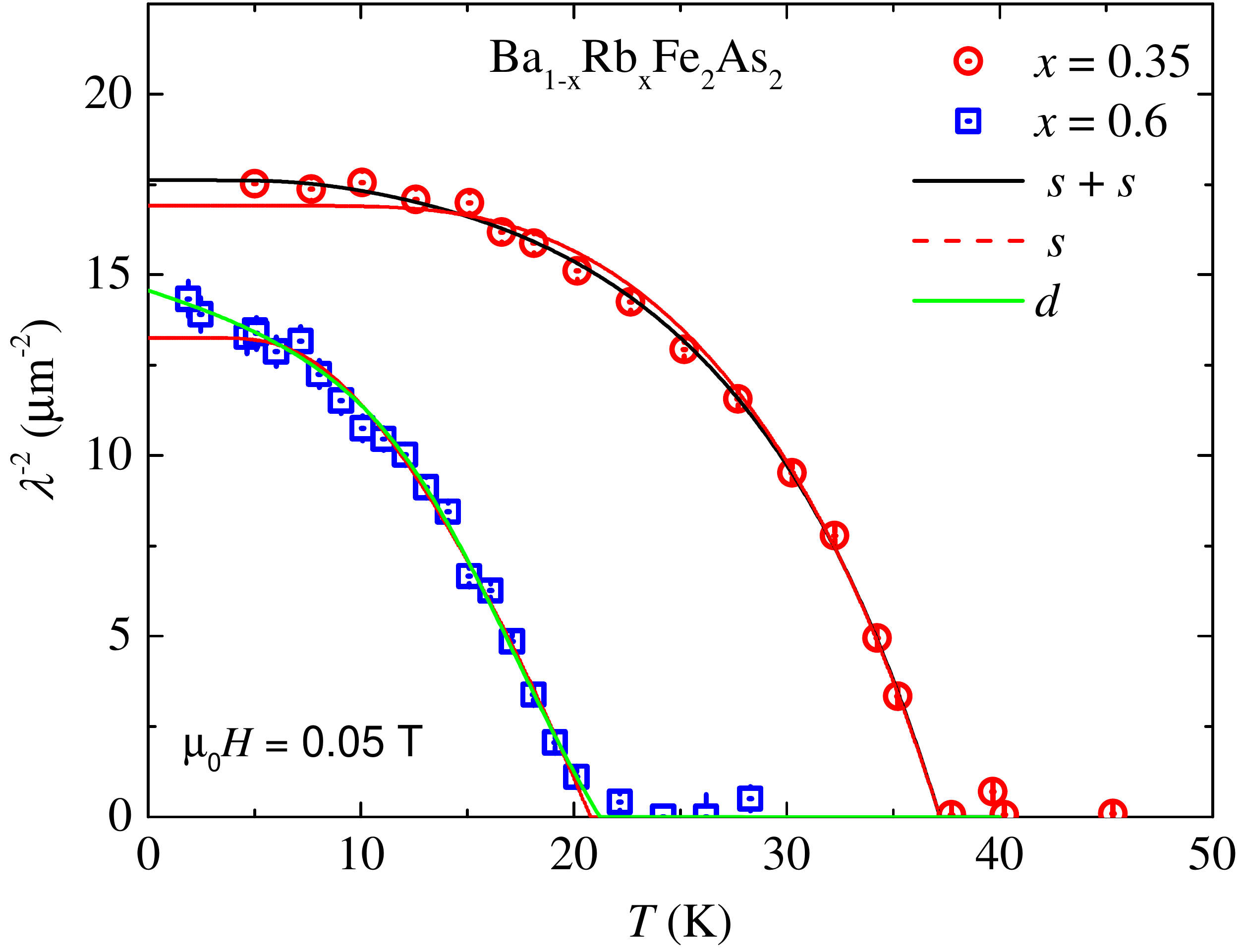}
\vspace{-0.5cm}
\caption{ (Color online) The temperature dependence of ${\lambda}^{-2}$ for
Ba$_{1-x}$Rb$_{x}$Fe$_{2}$As$_{2}$ ($x$ = 0.35, 0.65), measured in an applied field of
${\mu}_{\rm 0}H=0.05$~T. The dashed lines correspond
to a single gap BCS $s$-wave model, whereas the solid ones represent fits using a two-gap ($s+s$)-wave and a $d$-wave models.}
\label{fig3}
\end{figure}

\begin{equation}
\begin{aligned}
A_{ZF}(t)=V_{m}{A_0}\Bigg[\frac{1}{3}e^{-\lambda_{L}t}+\frac{2}{3}e^{-\lambda_{T}t}\Bigg]\\
+(1-V_{m}){A_0}\Bigg[\frac{1}{3}+\frac{2}{3}(1-{\sigma}^{2}t^{2}-{\Lambda}t)e^{{(-\frac{{\sigma^2}t^2}{2}}-{\Lambda}t)}\Bigg].
\label{eq1}
\end{aligned}
\end{equation}
Here, the first and the second terms describe the magnetic and non-magnetic parts of the signals, respectively.
$A_{0}$ is the initial asymmetry, $V_{\rm m}$ is the magnetic volume fraction, and the ${\lambda}_{\rm T}$, ${\lambda}_{\rm L}$
are the transverse and longitudinal depolarization rates of the ${\mu}$SR signal, respectively, arising from the magnetic part of the sample.
The second term describing the paramagnetic part of the sample is the combination of Lorentzian and Gaussian Kubo-Toyabe depolarization function \cite{Kubo,Hayano}. The depolarization rates  ${\sigma}$ and ${\Lambda}$ are due to the nuclear dipole moments and randomly oriented diluted local electronic moments, respectively. The magnetic fraction obtained by fitting Eq.~1 to the data for Ba$_{0.65}$Rb$_{0.35}$Fe$_{2}$As$_{2}$ and Ba$_{0.35}$Rb$_{0.65}$Fe$_{2}$As$_{2}$ is plotted in Fig.~4(c). The magnetic fraction $V_{\rm m}$ was found to be only 10 ${\%}$ in both samples at low temperatures, it decreases upon increasing the temperature and becomes negligibly small at around 80 K. 
By using the Eq.~1 to extract the magnetic fraction we assume that the static magnetic order causes the initial loss of asymmetry. If the loss of asymmetry is caused by dynamic magnetism, than the estimated magnetic fraction will be even lower by an overall scale factor as compared to the one shown in Fig. 4c. We note the non-monotonous temperature dependence of the magnetic fraction for Ba$_{0.35}$Rb$_{0.65}$Fe$_{2}$As$_{2}$.  $V_{\rm m}$ increases below 75 K. This kind of $T$-dependence of $V_{\rm m}$ was previously observed in a number of Fe-HTS's \cite{Bendele,Goltz} with the static magnetism and may be caused by the interplay between magnetism and superconductivity.


 The TF ${\mu}$SR data were analyzed by using the following functional form:\cite{Bastian,Khasanov,Yaouance}
\begin{equation}
A_{TF_S}(t)=A_Se^{-{\Lambda_{TF}}t}e^{\Big[-\frac{(\sigma_{sc}^2+\sigma_{nm}^2)t^2}{2}\Big]}\cos(\gamma_{\mu}B_{int}t+\varphi), 
\label{eq1}
\end{equation}
Here $A$ denotes the initial asymmetry, $\gamma/(2{\pi})\simeq 135.5$~MHz/T 
is the muon gyromagnetic ratio, and ${\varphi}$ is the initial phase of the muon-spin ensemble.
${\Lambda}$ is the exponential relaxation rate caused by the presence
of diluted Fe moments. $B_{\rm int}$ represents the
internal magnetic field at the muon site, and the relaxation rates ${\sigma}_{\rm sc}$ 
and ${\sigma}_{\rm nm}$ characterize the damping due to the formation of the flux-line lattice (FLL) in the SC state and of the nuclear magnetic dipolar contribution, respectively. During the analysis ${\sigma}_{\rm nm}$ was assumed 
to be constant over the entire temperature range and was fixed to the value obtained above 
$T_{\rm c}$ where only nuclear magnetic moments contribute to the muon depolarization rate ${\sigma}$.
Note that the Eq. 2 has been used previously for Fe-HTS`s in the presence of the diluted Fe-moments and it  
was demonstrated to be precise enough to extract the SC depolarisation rate as a function of temperature \cite{Khasanov}.
The temperature dependence of the difference between the internal field ${\mu}_{\rm 0}$$H_{\rm int,SC}$ measured in SC state 
and one ${\mu}_{\rm 0}$$H_{\rm int,NS}$ measured in the normal state at $T$ = 45 K for 
Ba$_{1-x}$Rb$_{x}$Fe$_{2}$As$_{2}$  ($x$ = 0.35 and 0.65) is shown in Fig.~5a, revealing a strong diamagnetic shift imposed by the SC state.
In Fig.~5b ${\sigma}_{\rm sc}$ is plotted as a function of temperature for 
Ba$_{1-x}$Rb$_{x}$Fe$_{2}$As$_{2}$  ($x$ = 0.35 and 0.65) 
at ${\mu}_{\rm 0}H=0.05$~T. Below $T_{\rm c}$ the relaxation rate ${\sigma}_{\rm sc}$ starts to increase from zero due to the
formation of the FLL. The value of ${\sigma}_{\rm sc}$ is lower for the over-doped sample than the one of the optimally-doped system. In addition, an interesting experimental fact is that the $T$-dependence of the relaxation rate, which reflects the topology of the SC gap, changes between $x$ = 0.35 and 0.65. The data for $x$ = 0.35 is flat below $T$/$T_{\rm c}$ ${\simeq}$ 0.4, indicating a fully gapped SC state. In contrast, the data for $x$ = 0.65 exhibits a steeper temperature dependence of ${\sigma}_{\rm sc}$($T$), indicating the presence of quasiparticle excitations.
In order to quantify the change of the symmetry of the SC gap, we analyzed the $T$-dependence of the magnetic penetration depth.
For polycrystalline samples the temperature dependence of the London magnetic penetration depth 
${\lambda}(T)$ is related to the muon spin depolarization rate ${\sigma}_{\rm sc}(T)$ by the equation:\cite{Brandt}
\begin{equation}
\frac{\sigma_{sc}^2(T)}{\gamma_\mu^2}=0.00371\frac{\Phi_0^2}{\lambda^4(T)},
\end{equation}          
where ${\Phi}_{\rm 0}=2.068 {\times} 10^{-15}$~Wb is the magnetic-flux quantum. 
Equation (2) is only valid, when the separation between the vortices is smaller than ${\lambda}$. In this case according
to the London model ${\sigma}_{\rm sc}$ is field independent.\cite{Brandt}
Field dependent measurements of ${\sigma}_{\rm sc}$
was reported previously \cite{GuguchiaN}. It was observed that first
${\sigma}_{\rm sc}$ strongly increases with increasing magnetic field until reaching a maximum
at ${\mu}_{\rm 0}H$  ${\simeq}$  0.03~T and then above 0.03~T stays nearly constant up to the highest field 
(0.64~T) investigated. Such a behavior is expected within the London model and is typical for polycrystalline
high temperature superconductors (HTS's) \cite{Pumpin}. The observed field dependence of ${\sigma}_{\rm sc}$ implies that for a reliable determination of
the penetration depth the applied field must be larger than ${\mu}_{\rm 0}H= 0.03$~T.

\begin{figure}[b!]
\includegraphics[width=1.0\linewidth]{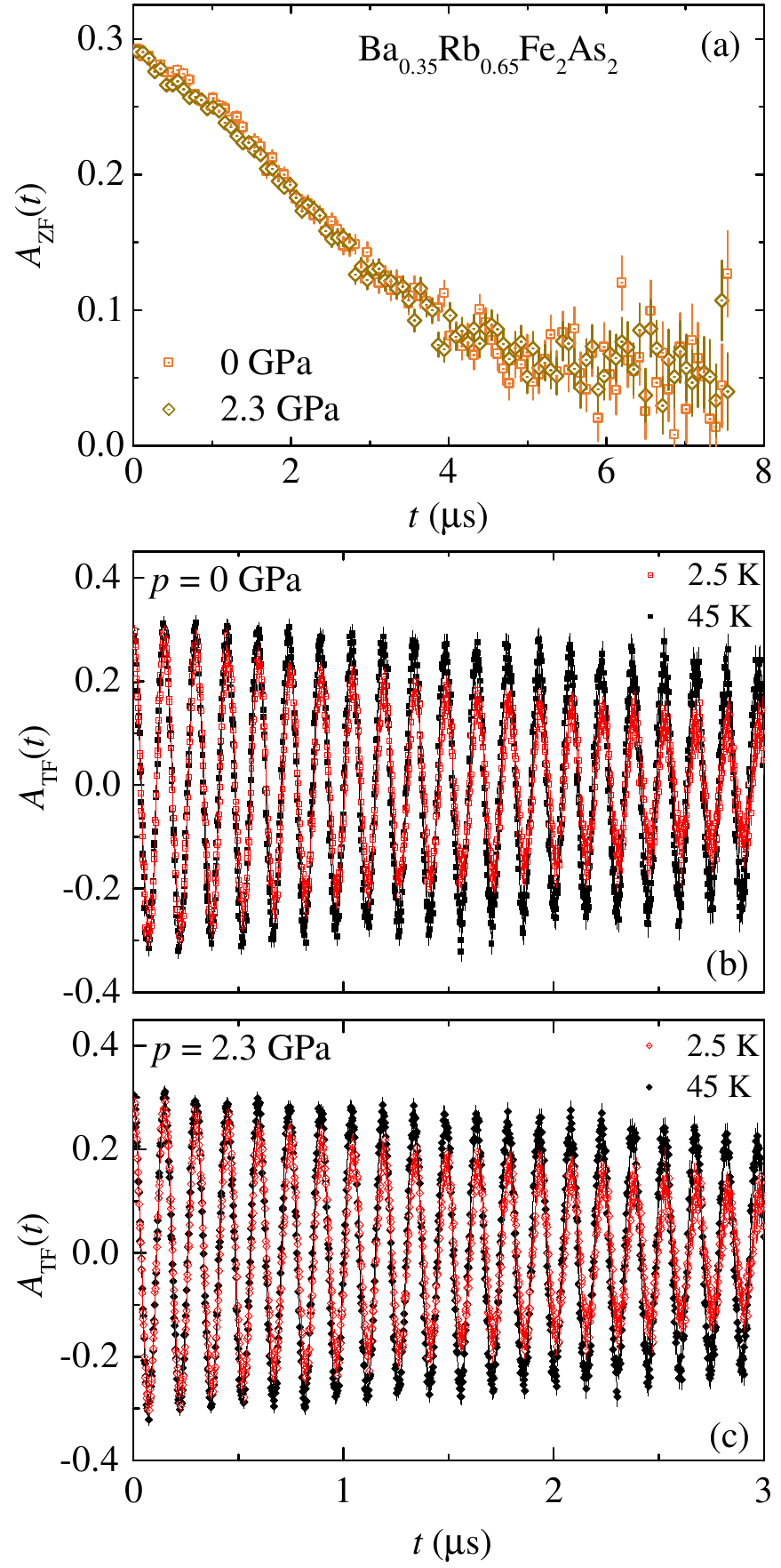}
\vspace{-0.5cm}
\caption{ (Color online) (a) ZF-${\mu}$SR time spectra for Ba$_{0.35}$Rb$_{0.65}$Fe$_{2}$As$_{2}$ for
$p$ = 0 and 2.3 GPa at the base temperature $T$ = 2.5 K. Transverse-field (TF) ${\mu}$SR time spectra obtained above and below $T_{\rm c}$ for Ba$_{0.35}$Rb$_{0.65}$Fe$_{2}$As$_{2}$ (after field cooling the sample from above $T_{\rm c}$): (b)  $p$ = 0 GPa and (c) $p$ = 2.3 GPa. The solid lines in panel (b) and (c) represent fits to the data by means of Eq.~6.}
\label{fig3}
\end{figure}
\begin{figure}[b!]
\includegraphics[width=1.0\linewidth]{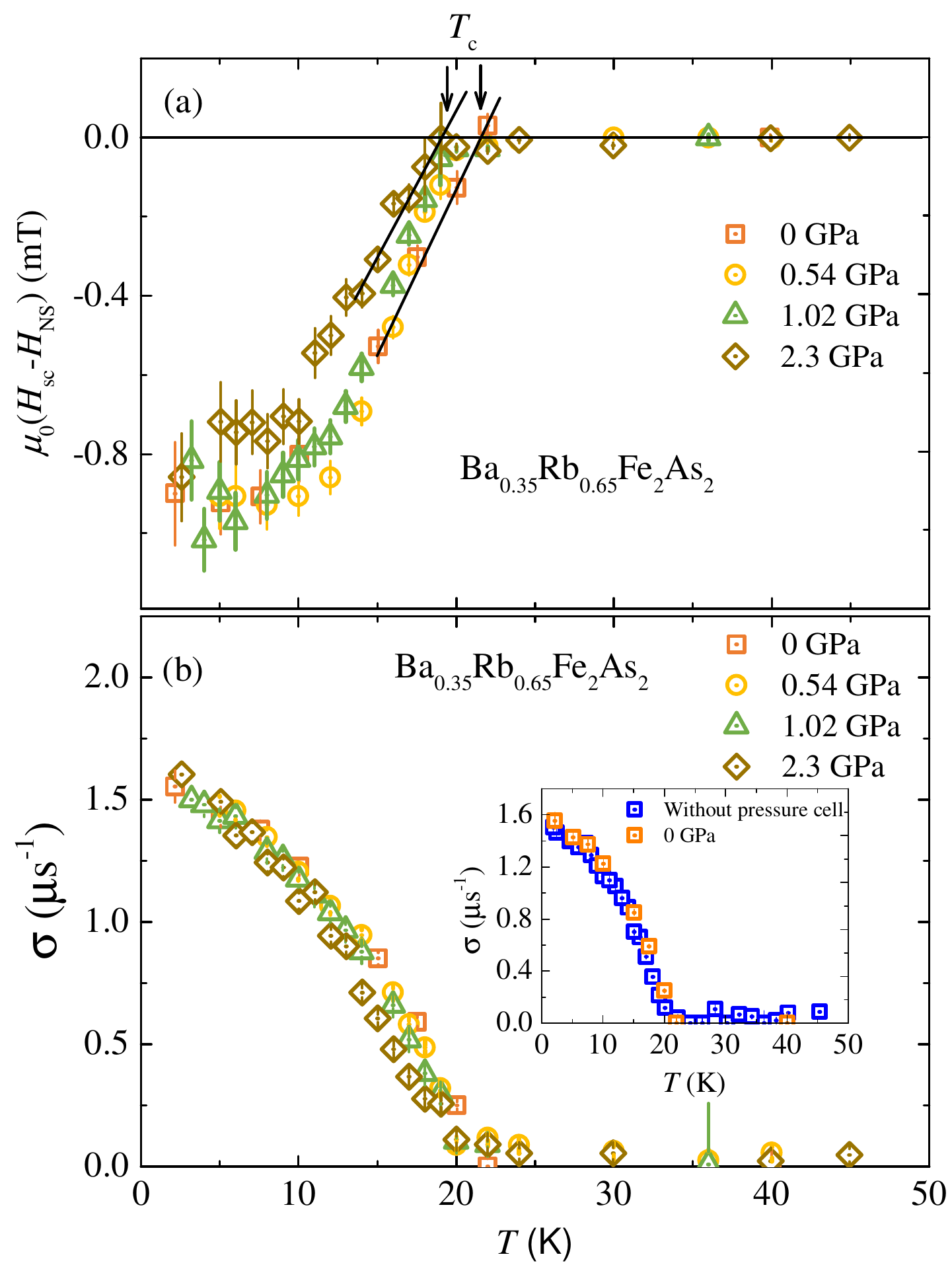}
\vspace{-0.5cm}
\caption{ (Color online) (a) Temperature dependence of the difference
between the internal field ${\mu}_{\rm 0}$$H_{\rm SC}$ measured in the SC state 
and the one measured in the normal state ${\mu}_{\rm 0}$$H_{\rm NS}$ at $T$ = 45~K  for Ba$_{0.65}$Rb$_{0.35}$Fe$_{2}$As$_{2}$ 
recorded for various hydrostatic pressures. (b) Temperature dependence of the superconducting muon spin depolarization rate
${\sigma}_{\rm sc}$ in an applied magnetic field of ${\mu}_{\rm 0}H = 50$~mT for   
Ba$_{0.35}$Rb$_{0.65}$Fe$_{2}$As$_{2}$ for selected applied pressures. The inset shows the data recorded at $p$ = 0 GPa for the sample measured together with the cell and without the cell.}
\label{fig3}
\end{figure}

 ${\lambda}$($T$) can be calculated within the local (London) approximation (${\lambda}$ ${\gg}$ ${\xi}$) 
by the following expression:\cite{Bastian,Tinkham}
\begin{equation}
\frac{\lambda^{-2}(T,\Delta_{0,i})}{\lambda^{-2}(0,\Delta_{0,i})}=
1+\frac{1}{\pi}\int_{0}^{2\pi}\int_{\Delta(_{T,\varphi})}^{\infty}(\frac{\partial f}{\partial E})\frac{EdEd\varphi}{\sqrt{E^2-\Delta_i(T,\varphi)^2}},
\end{equation}
where $f=[1+\exp(E/k_{\rm B}T)]^{-1}$ is the Fermi function, ${\varphi}$ is the angle along the Fermi surface, and ${\Delta}_{i}(T,{\varphi})={\Delta}_{0,i}{\Gamma}(T/T_{\rm c})g({\varphi}$)
(${\Delta}_{0,i}$ is the maximum gap value at $T=0$). 
The temperature dependence of the gap is approximated by the expression 
${\Gamma}(T/T_{\rm c})=\tanh{\{}1.82[1.018(T_{\rm c}/T-1)]^{0.51}{\}}$,\cite{carrington} 
while $g({\varphi}$) describes 
the angular dependence of the gap and it is replaced by 1 for both an $s$-wave and an $s$+$s$-wave gap,
and ${\mid}\cos(2{\varphi}){\mid}$ for a $d$-wave gap.\cite{Fang}   

 The temperature dependence of the penetration depth was analyzed using either a single gap or a two-gap model 
which is based on the so-called ${\alpha}$ model. This model was first discussed by 
Padamsee $et$ $al$.\cite{padamsee} and later on was succesfully used to analyse the magnetic penetration depth data in HTS's \cite{carrington,khasanovalpha}.
According to the ${\alpha}$ model, the superfluid density is calculated for each component using Eq.~3
and then the contributions from the two components added together, $i.e.$,  
\begin{equation}
\frac{\lambda^{-2}(T)}{\lambda^{-2}(0)}=\omega_1\frac{\lambda^{-2}(T,\Delta_{0,1})}{\lambda^{-2}(0,\Delta_{0,1})}+\omega_2\frac{\lambda^{-2}(T,\Delta_{0,2})}{\lambda^{-2}(0,\Delta_{0,2})},
\end{equation}\\
where ${\lambda^{-2}}(0)$ is the penetration depth at zero temperature, ${\Delta_{0,i}}$ is the 
value of the $i$th ($i=1$, 2) superconducting gap at $T=0$~K, and ${\omega}_{i}$ is a weighting 
factor which measures their relative contributions to ${\lambda^{-2}}$ (${\omega}_{1}+{\omega}_{2}=1$). 

 The results of the analysis for Ba$_{1-x}$Rb$_{x}$Fe$_{2}$As$_{2}$ ($x=0.35$, 0.65) 
are presented in Fig.~6. The lines represent fits to the data using
a $s$-wave  model (dashed line) and $s+s$-wave and $d$-wave models (solid lines). 
In agreement with our previous report \cite{GuguchiaN} the two-gap 
$s$+$s$-wave scenario with a small gap ${\Delta}_{1}$ = 2.7(5) meV and a large gap
${\Delta}_{2}$ = 8.41(23) meV, describes the experimental data for the optimally doped $x$ = 0.35 sample fairly well.
On the other hand, for the over-doped sample $x$ = 0.65 a $d$-wave gap symmetry with a gap ${\Delta}$ = 8.2(7) meV gives an  adequate description of ${\lambda^{-2}}(T)$. This conclusion is supported by a ${\chi}^{2}$ test, revealing the smaller value of ${\chi}^{2}$ for $d$-wave model by ${\sim}$ 25 ${\%}$  than the one for  $s+s$-wave model. 
Furthermore, we note that the $s+s$-wave model fit gives the following value for the smaller gap ${\Delta}_{1}$ = 0.6(5) meV which is comparable to zero, ruling out this model  as a possible description of ${\lambda}(T)$ for the over-doped sample $x$ = 0.65.
This suggests that the heavy hole-doping in Ba$_{1-x}$Rb$_{x}$Fe$_{2}$As$_{2}$ induces a change of the SC gap topology from nodless to nodal, as it is the case for the related system  Ba$_{1-x}$K$_{x}$Fe$_{2}$As$_{2}$.


\subsubsection{High pressure zero-field and transverse-field ${\mu}$SR experiments on Ba$_{0.35}$Rb$_{0.65}$Fe$_{2}$As$_{2}$}
\begin{figure*}[t!]
\includegraphics[width=1\linewidth]{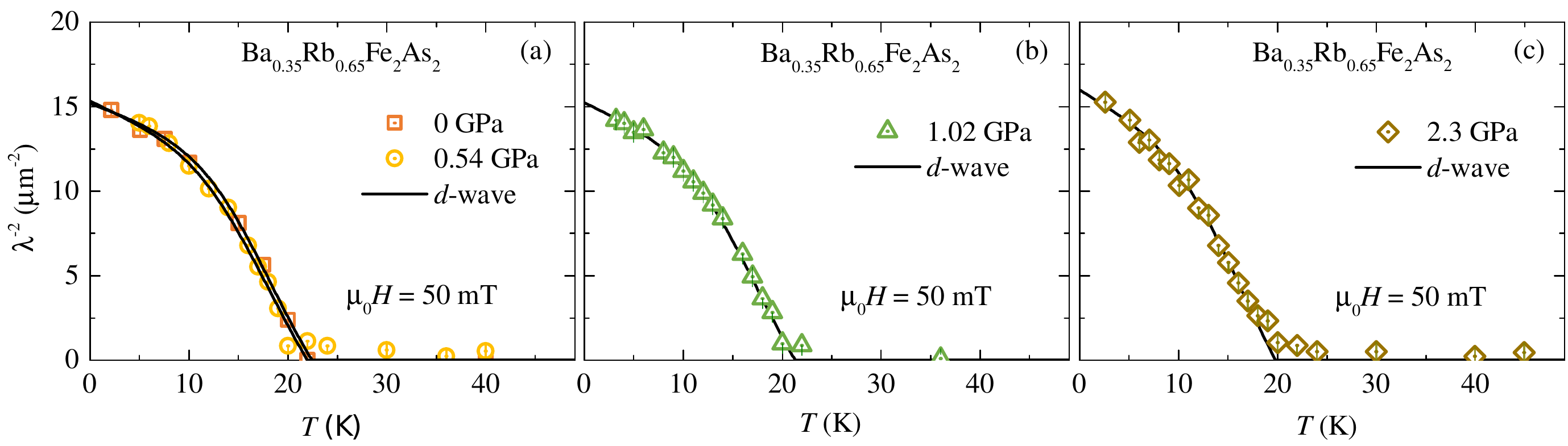}
\vspace{-0.5cm}
\caption{ (Color online) (a-c) The temperature dependence of ${\lambda}^{-2}$ 
measured at various applied hydrostatic pressures for Ba$_{0.35}$Rb$_{0.65}$Fe$_{2}$As$_{2}$ in an applied field of
${\mu}_{\rm 0}H=50 $~mT. The solid lines correspond to a single-gap $d$-wave model.}
\label{fig3}
\end{figure*}
Figure~7a shows the ZF-${\mu}$SR time spectra for $p$ = 0 and 2.3 GPa obtained at  $T$ = 2.5 K for 
Ba$_{0.35}$Rb$_{0.65}$Fe$_{2}$As$_{2}$.
The ZF relaxation rate stays nearly unchanged between $p$ = 0 GPa and 2.3 GPa, implying that there is no sign of pressure induced magnetism in this system.  
Figures~7b and 7c exhibit the TF time spectra for Ba$_{0.65}$Rb$_{0.35}$Fe$_{2}$As$_{2}$, measured at ambient $p$ = 0 GPa and maximum applied pressure $p$ = 2.22 GPa, respectively. Spectra above (45 K) and below (1.7 K) the SC transition temperature 
$T_{\rm c}$ are shown.
 The TF ${\mu}$SR data were analyzed by using the following functional form:\cite{Bastian}
\begin{equation}
\begin{aligned}
A_{TF}(t) = A_{TF_S}(t) + A_{pc}\exp\Big[-\frac{\sigma_{pc}^2t^2}{2}\Big]\cos(\gamma_{\mu}B_{int,pc}t+\varphi). 
\end{aligned}
\end{equation}
 Here $P_{\rm s}(t)$ is the function used to describe the sample response and is given by Eq.~2. $A_{\rm pc}$  denote the initial asymmetry of the pressure cell. ${\varphi}$ is the initial phase of the muon-spin ensemble and $B_{\rm int,pc}$ represents the internal magnetic field probed by the muons, stopped in the pressure cell. 
The Gaussian relaxation rate, ${\sigma}_{\rm pc}$, reflects the depolarization due
to the nuclear magnetism of the pressure cell.  
As shown previously \cite{Maisuradze-PC}, the diamagnetism of the SC sample has an influence on the pressure cell signal, leading to the temperature dependent ${\sigma}_{\rm pc}$ below $T_{c}$. In order to consider the influence of the diamagnetic moment of the sample on the pressure cell \cite{Maisuradze-PC} we assume the
linear coupling between ${\sigma}_{\rm pc}$ and the field shift of the internal magnetic field in the SC state: 
${\sigma}_{\rm pc}$($T$) = ${\sigma}_{\rm pc}$($T$ ${\textgreater}$ $T_{\rm c}$) + $c(T)$(${\mu}_{\rm 0}$$H_{\rm int,NS}$ - ${\mu}_{\rm 0}$$H_{\rm int,SC}$), where  ${\sigma}_{\rm pc}$($T$ ${\textgreater}$ $T_{\rm c}$) = 0.35 ${\mu}$$s^{-1}$ is the temperature independent Gaussian relaxation rate. 
${\mu}_{\rm 0}$$H_{\rm int,NS}$ and ${\mu}_{\rm 0}$$H_{\rm int,SC}$ are the internal magnetic fields measured in the normal and in the SC state, respectively. As indicated by the solid lines in Figs.~7(b,c), the ${\mu}$SR data are well described by Eq.~(5).

 A large diamagnetic shift of ${\mu}_{\rm 0}$$H_{\rm int}$ sensed by the muons below $T_{\rm c}$ is observed at all applied pressures.
This is evident in Fig.~8a, where we plot the difference between the internal field ${\mu}_{\rm 0}$$H_{\rm int,SC}$ measured in SC state 
and ${\mu}_{\rm 0}$$H_{\rm int,NS}$ measured in the normal state at  $T$ = 45 K for Ba$_{0.35}$Rb$_{0.65}$Fe$_{2}$As$_{2}$. 
The SC transition temperature $T_{\rm c}$ is determined from the intercept of the   
linearly extrapolated ${\mu}_{\rm 0}$($H_{\rm int,SC}$-$H_{\rm int,NS}$) curve with it's zero line and it is found to be 
$T_{\rm c}$ = 21.6(7) K and 19.2(5) for $p$ = 0 GPa and 2.3 GPa, respectively. The ambient pressure value of $T_{\rm c}$
is in perfect agreement with the one $T_{\rm c}$ = 20.9(5) K obtained from magnetization and specific heat experiments.
With the highest pressure applied $p$ = 2.3 GPa $T_{\rm c}$  decreases by ${\sim}$ 2.4 K, 
corresponding to a stronger pressure effect on  $T_{\rm c}$ as compared to the one observed in the optimally doped sample Ba$_{0.65}$Rb$_{0.35}$Fe$_{2}$As$_{2}$. 
The temperature dependence of ${\sigma}_{\rm sc}$ for Ba$_{0.35}$Rb$_{0.65}$Fe$_{2}$As$_{2}$ at various pressures is shown in Fig.~8b.
The inset shows the data recorded at $p$ = 0 GPa for the sample measured together with the cell and without the cell. The temperature dependences as well as the relaxation rates are in good agreement with each other. 
While the overlap of the low-$T$ data for $p$ = 0, 0.54 and 1.02 GPa is observed, it is clear that ${\sigma}_{\rm sc}$ for the highest applied pressure $p$ = 2.3 GPa has slightly steeper $T$-dependence at low temperatures. This leads to a tiny pressure induced decrease of the zero-temperature value of the penetration depth ${\lambda}$(0), 
which is different from the observation for the $x$ = 0.35 sample where a substantial decrease of ${\lambda}$(0) was reported. Note that for all applied pressures the temperature dependence of ${\lambda}^{-2}$ is well described by a $d$-wave gap symmetry as shown in Fig.~9. This implies that the $d$-wave symmetry in over-doped Ba$_{0.35}$Rb$_{0.65}$Fe$_{2}$As$_{2}$ is robust against pressure. 
The results of the $x$ = 0.65 sample extracted from the analysis of the pressure data are summarized in Table I.
We note that the nodal SC gaps are promoted in the optimally doped $x$ = 0.35 system under pressure, as shown in our previous work \cite{GuguchiaN}. However, the nodes exist only on the electron pockets, while in the hole pocket gap is nearly constant.
But in case of the over-doped $x$ = 0.65 system, the results are consistent with the presence of nodal $d$-wave gaps on all Fermi surface sheets.
It is important to emphasize that by heavy hole doping as well as hydrostatic pressure,
one can induce stable $d$-wave pairing in Ba$_{1-x}$Rb$_{x}$Fe$_{2}$As$_{2}$.  
The recent  theoretical and experimental \cite{Bohm} studies of optimally-doped Ba$_{0.6}$K$_{0.4}$Fe$_{2}$As$_{2}$ revealed a sub-dominant $d$-wave state close in energy to an $s^{+- }$ state. 
It was shown that the coupling strength in this subdominant $d$ channel is as strong as 60 ${\%}$ of that in the dominant $s^{+- }$ channel. According to the results, presented and discussed above, pressure and heavy hole doping tip the intricate balance between $d$ and $s$ in favor of a $d$-wave state.

\begin{table}[t!]
\caption{Summary of the parameters obtained for polycrystalline samples of  Ba$_{0.35}$Rb$_{0.65}$Fe$_{2}$As$_{2}$ by means of ${\mu}$SR.}
\vspace{0.3cm}
\begin{tabular}{lcccc}
\hline
\hline
        $p$ (GPa)                                       &  $T_{\rm c}$ (K)    & ${\Delta}$ (meV)  & 2${\Delta}/k_{\rm B}T_{\rm c}$ & ${\lambda}$ (nm) \\ \hline 
0                               &  22.44(13)  & 8.2(7) & 8.55(47)  & 257(5)\\ 
 0.54                           &  21.97(15) & 6.4(3) & 6.83(47) & 254(5) \\
1.02                               & 21.32(18) & 6.3(3) & 6.88(59) & 256(6)\\ 
2.3                            &  19.8(2) & 5.3(3) & 6.1(6) & 250(6) \\ 
\hline         
\end{tabular}
\label{table1}
\end{table}
\begin{figure*}[t!]
\includegraphics[width=0.95\linewidth]{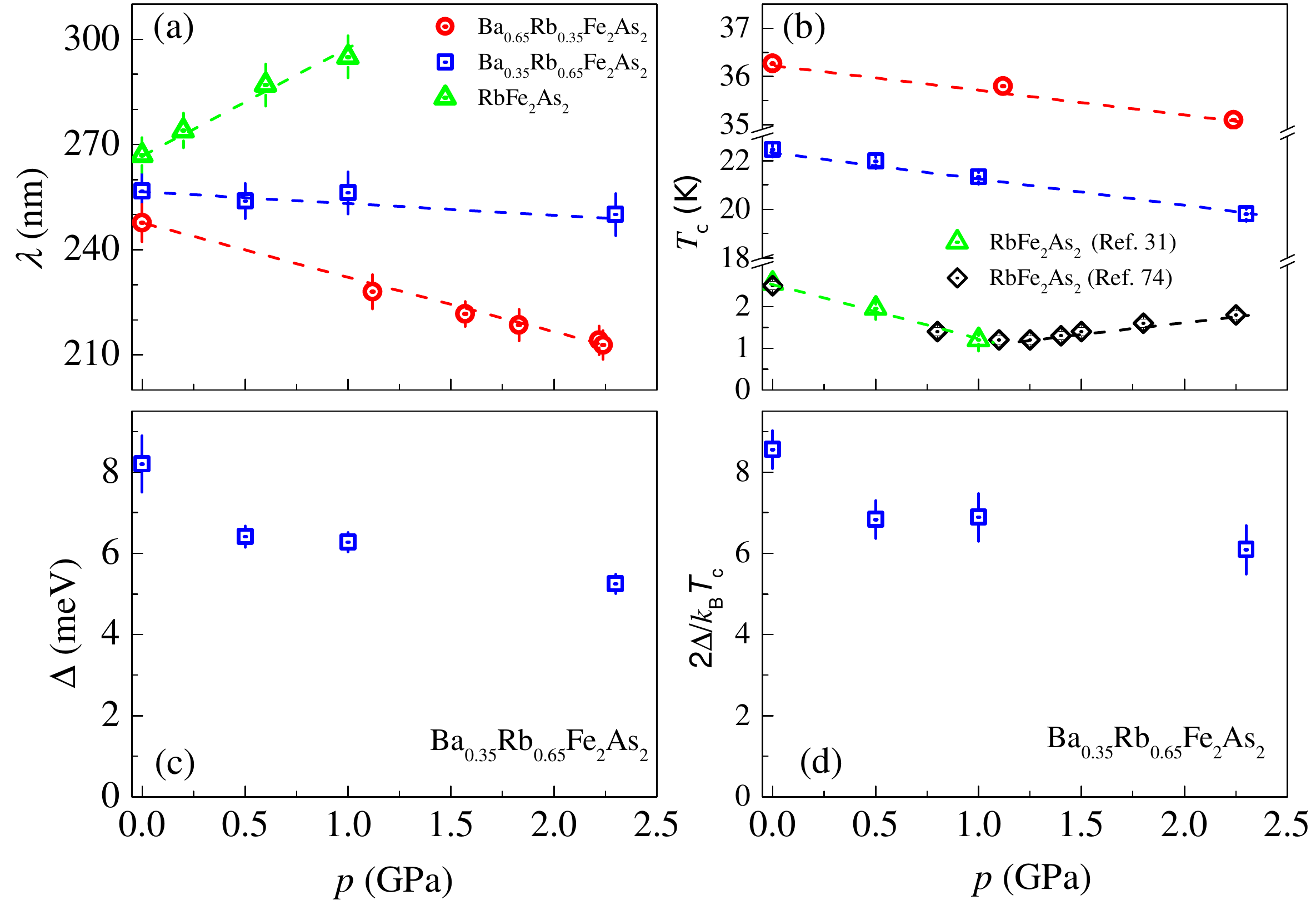}
\caption{ (Color online) The zero temperture value of the magnetic penetration depth ${\lambda}$(0) (a) and 
the SC transition temperature $T_{\rm c}$ (b) for Ba$_{1-x}$Rb$_{x}$Fe$_{2}$As$_{2}$ ($x$ = 0.35, 0.65, 1.0), as well as the $d$-wave gap (c) and the gap to $T_{c}$ ratio 2${\Delta}/k_{\rm B}T_{\rm c}$ (d) for the over doped sample
$x$ = 0.65, plotted as function of hydrostatic pressure. 
The measurements were performed in an applied magnetic field of ${\mu}_{\rm 0}H = 50$~mT. 
The data for $x$ = 0.35 sample are taken from Ref.~39 and the data for $x$ = 1 are taken from Ref.~31 and Ref.~74. The dashed lines represent the guides to the eyes.}
\label{fig4}
\end{figure*}

\section{PHASE DIAGRAM}

 The results of ${\lambda}^{-2}$($T$) for  Ba$_{0.35}$Rb$_{0.65}$Fe$_{2}$As$_{2}$ analysis are summarized in Fig.~10a-d, showing $T_{\rm c}$ as well as the zero-temperature values of ${\lambda}$(0), the SC $d$-wave gap ${\Delta}$, and the gap to $T_{\rm c}$ ratio 2${\Delta}$/$k_{B}$$T_{\rm c}$
as a function of hydrostatic pressure. Upon increasing the hydrostatic pressure from $p$ = 0 to 2.3 GPa, ${\lambda}$(0) is increased by less than 5 ${\%}$ and  $T_{\rm c}$ is decreased by 10 ${\%}$. Both ${\Delta}$ and 2${\Delta}$/$k_{B}$$T_{\rm c}$ shows only a modest decrease with increasing pressure.
Our results show that there are no significant changes of the SC properties of Ba$_{0.35}$Rb$_{0.65}$Fe$_{2}$As$_{2}$ under pressure and
$d$ represents the most stable pairing symmetry in Ba$_{0.35}$Rb$_{0.65}$Fe$_{2}$As$_{2}$.  
In order to reach a more complete view of the pressure effect on ${\sigma}_{\rm sc}$(0) and $T_{\rm c}$  in Ba$_{1-x}$Rb$_{x}$Fe$_{2}$As$_{2}$ in Fig.~10a,b we combined the present data with the previous high-pressure ${\mu}$SR results on optimally doped Ba$_{0.65}$Rb$_{0.35}$Fe$_{2}$As$_{2}$ and on RbFe$_{2}$As$_{2}$ which presents the case of a naturally over-doped system. 
For all samples, the $T_{\rm c}$($p$) and ${\lambda}$(0)($p$) behaviors are linear, so that the pressure dependence of $T_{\rm c}$ and ${\lambda}$(0) can be well represented by  $dT_{\rm c}$/$dp$ and $d{\lambda}$(0)/$dp$ values, respectively. The pressure derivative, $dT_{\rm c}$/$dp$, is negative for all $x$ = 0.35, 0.65, 1 and it's magnitude increases with increasing $x$. However, there is at higher pressures  fundamental difference of $T_{\rm c}$($p$) between the $x$ = 0.35, 0.65 and $x$ = 1 samples. Namely, for $x$ = 1 a $V$-shaped temperature pressure phase diagram is observed \cite{TailleferPRB} as in KFe$_{\rm 2}$As$_{2}$ \cite{Taillefer} which is absent for  $x$ = 0.35 and 0.65. Regarding  ${\lambda}$(0), application of pressure of $p$ = 2.3 GPa causes a decrease of its value by 15 ${\%}$ in optimally doped sample $x$ = 0.35, while only very tiny decrease of ${\lambda}$(0) is observed for the over-doped system $x$ = 0.65. Instead, for the end member compound an increase of ${\lambda}$(0) with pressure is observed. This means that the $d{\lambda}$(0)/$dp$ is negative and large for $x$ = 0.35. On further increasing the $x$ to 0.65 it's magnitude becomes negligibly small but is still negative and becomes positive for the end member $x$ = 1. So, the sign change of $d{\lambda}$(0)/$dp$ takes place for some the $x$ values located between $x$ = 0.65 and 1. The above results provide clear evidence that the SC gap symmetry as well as the pressure effects on $T_{\rm c}$ and on ${\lambda}$(0) strongly depends on doping level $x$. 
Note that in the optimally doped '122'-system Ba$_{1-x}$K$_{x}$Fe$_{2}$As$_{2}$ several bands cross the Fermi surface (FS) \cite{Evtushinsky,Ding,Zabolotnyy}.  They consist of inner (${\alpha}$) and outer (${\beta}$) hole-like bands, 
both centered at the zone center ${\Gamma}$, and an electron-like band (${\gamma}$) centered at the M point.
Band structure of Ba$_{1-x}$K$_{x}$Fe$_{2}$As$_{2}$ changes are associated with hole doping.
The hole Fermi surfaces expand with increasing $x$, whereas electron Fermi surfaces shrink gradually and disappear
for $x$ ${\textgreater}$ 0.6, giving rise to a Lifshitz transition. Since, the investigated system is very similar to 
Ba$_{1-x}$K$_{x}$Fe$_{2}$As$_{2}$, one expects similar doping induced changes in the band structure in both materials. Hence, the $x$ dependence of the SC gap symmetry as well as the pressure effects, reported above for Ba$_{1-x}$Rb$_{x}$Fe$_{2}$As$_{2}$, may be related to this putative Lifshitz transition.

\section{SUMMARY AND CONCLUSIONS}

 In summary, the SC properties of  optimally and over-doped Ba$_{1-x}$Rb$_{x}$Fe$_{2}$As$_{2}$ ($x=0.35$ and 0.65) samples at ambient pressure were studied by means of magnetization, specific heat and ${\mu}$SR experiments.
In addition, the $x$ = 0.65 specimen was investigated under hydrostatic pressures up to $p$ = 2.3 GPa through
zero-field and transverse field ${\mu}$SR experiments.
While the specific heat jump for the $x$ = 0.35 sample follows the so called BNC scalling, 
the heavily over-doped  Ba$_{1-x}$Rb$_{x}$Fe$_{2}$As$_{2}$ shows a 
deviation from the BNC scaling as it was observed for the related Ba$_{1-x}$K$_{x}$Fe$_{2}$As$_{2}$ system.
In contrast to nodeless SC gap observed in the optimally doped sample $x$ = 0.35, the temperature dependence of the magnetic penetration depth ${\lambda}$ suggests a $d$-wave SC gap in over-doped system $x$ = 0.65.
The $d$-wave symmetry is preserved under hydrostatic pressures up to
$p$ = 2.3 GPa, indicating the robustness of the $d$-wave symmetry in the over-doped region.
The fact that the rather stable $d$-wave symmetry was also observed in 
the optimally-doped sample $x$ = 0.35 under pressure
indicates that both tuning parameters, heavy hole doping and hydrostatic pressure, promote the same pairing mechanism for superconductivity in
 $Ba_{1-x}$Rb$_{x}$Fe$_{2}$As$_{2}$.
The values of the magnetic penetration depth ${\lambda}$, $T_{\rm c}$ as well as the
$d$-wave gap ${\Delta}$ and the ratio $2{\Delta}/k_{\rm B}T_{\rm c}$ show a small and monotonic decrease with 
increasing the pressure.
By combining the present data with those previously obtained for the optimally doped system \cite{GuguchiaSubmitted} and for the end member RbFe$_{2}$As$_{2}$ \cite{Shermadinipressure} we conclude 
that the SC gap symmetry as well as the pressure effects on the quantities characterizing the SC state strongly depends on the hole doping level $x$.
The combined results may be interpreted by assuming a disappearance of the electron pocket from the Fermi surface upon the high hole doping, resulting in
a Lifshitz transition. Note that the absence of the ${\gamma}$ electron pocket has been observed by ARPES in the related system KFe$_{2}$As$_{2}$ \cite{Sato}.
Finally, we suggest that the Ba$_{1-x}$Rb$_{x}$Fe$_{2}$As$_{2}$ and Ba$_{1-x}$K$_{x}$Fe$_{2}$As$_{2}$ superconducting series have a common doping dependence of the SC properties. The present results may help to explore the microscopic mechanism responsible for the observed non-universaltiy
of the SC gap structure in the Fe-HTS's.

\section{Acknowledgments}~
The work was performed at the Swiss Muon Source (S${\mu}$S)
Paul Scherrer Insitute, Villigen, Switzerland. 
Z.G. acknowledge the support by the Swiss National Science Foundation.

\end{document}